# SIGNLINE: Digital signature scheme based on linear equations cryptosystem

Gennady Khalimov [1[0000-0002-2054-9186]], Yevgen Kotukh [2[0000-0003-4997-620X]], Maksym Kolisnyk [1[0000-0002-1075-9470]], Svitlana Khalimova [1[0000-0001-7224-589X]], Oleksandr Sievierinov [1[0000-0002-6327-6405]]

[1] Kharkiv National University of Radioelectronics, Kharkiv, 61166, Ukraine
hennadii.khalimov@nure.ua
[2] Yevhenii Bereznyak Military Academy, Kyiv, Ukraine
yevgenkotukh@gmail.com

**Abstract.** The paper explores a novel cryptosystem for digital signatures based on linear equations for logarithmic signatures. A logarithmic signature serves as a fundamental cryptographic primitive, characterized by properties such as nonlinearity, non-commutability, unidirectionality, and key-dependent factorability. The proposed cryptosystem ensures the secrecy of logarithmic signatures through its foundation in linear equations. Quantum security is achieved by eliminating any possible mapping between the input and output of the logarithmic signature, thereby rendering Grover's quantum attack ineffective. The public key sizes for the NIST security levels of 128, 192, and 256 bits are 1, 1.5, and 2 KB, respectively. The algorithm demonstrates scalability concerning computational costs, memory usage, and hardware limitations without compromising security. Its primary operation involves bitwise XOR over logarithmic arrays of 8, 16, 32, and 64 bits.

## 1 Introduction

Since 2017, the National Institute of Standards and Technology (NIST) has initiated a competition to develop post-quantum cryptographic algorithms for public keys [1]. The NIST PQC standardization project aims to replace the current recommended digital signature standards, RSA and ECDSA, with post-quantum algorithms [2]. NIST received 69 submissions that met the minimum requirements for computational resources necessary to withstand successful attacks on key selection and collision searches across various security categories. From these submissions, 15 algorithms were selected for the third round: nine candidates for public-key encryption mechanisms (KEM) and six for digital signature schemes [3]. According to the results of this round, three digital signature algorithms emerged as finalists: Crystals-Dilithium (Dilithium), Falcon, and Rainbow (Dilithium [4], Falcon [5], and Rainbow [6]). These algorithms are grounded in design principles derived from lattice and multidimensional cryptography. The analysis of their security using classical computational methods is extensively documented [7].

Lattice-based signature schemes operate on a set of points in an n-dimensional space with a periodic structure [8]. The security of lattice-based cryptography is achieved through the use of NP-complex problems, such as the search for the shortest vectors (SVP, CVP, SVIP) and learning with errors (LWE, LWR) [3]. To ensure security, Dilithium relies on the Fiat-Shamir and Aborts structure, as well as SVP [4]. Falcon employs a compact signature based on a fast Fourier lattice over a ring of truncated polynomials of the Nth degree. For security, it utilizes NTRU for key generation, data encryption and decryption, short integer problems (SIS), floating-point arithmetic, and Gaussian floating-point sampling. The cryptanalysis of lattice-based algorithms is presented in [10] and [11]. The authors of [10] considered an attack involving lattice reduction based on additional information and hints, demonstrating how this approach can reduce the costs of performing attacks. In [11], attacks were analyzed where part of the secret key was extracted and used to forge signatures. Their analysis included miss-adding attacks on Dilithium and proposed solutions to mitigate them at zero cost.

Signatures based on multidimensional variables, known as unbalanced oil and vinegar (UOV) systems, involve hiding quadratic equations with $n$ unknowns in a finite field $k$ [12]. The security of these signature schemes is predicated on the complexity of solving quadratic equations over finite fields, making them NP-hard problems. The security of the scheme also depends on the number of variables and the size of the field, necessitating large key sizes. Rainbow, a multivariate signature scheme, ensures security by using a system of hidden quadratic equations over a binary field (LUOV) [13].

## 2  Motivation

Technological giants have initiated research in the field of quantum computing, with programs aimed at developing new quantum algorithms and demonstrating the advantages of quantum computers. Bart Prenell notes, "it should be noted that there are still no large quantum computers that could prove in a real way that quantum algorithms work." Currently, the leading candidates for NIST standards remain under consideration, but these are only preliminary impressions, and future studies may alter this perspective [14]. One of the earliest attempts to construct a polynomial quantum algorithm for solving the Learning With Errors (LWE) problem with polynomial modulus-noise relations failed [15]. Despite this, the author suggests that "new ideas on the application of the complex Gaussian function and the windowed quantum Fourier transform will be able to find new applications in quantum computing or develop new ways to solve the LWE problem" [16].

The application of NP-hard problems is pivotal in designing quantum-secure cryptographic systems. We posit that any cryptographic algorithm exhibiting regularity in its structured data is susceptible to being compromised by a quantum computer. The properties of superposition and quantum entanglement enable simultaneous calculations across all states of a qubit register, effectively modeling a complete set of classical computations. Regularity in the computational data of an algorithm, such as periodicity in algebraic structures (rings, groups, lattices, etc.), can potentially be exploited by a quantum algorithm with lower complexity than Grover's algorithm.

We propose a paradigm shift in cryptosystem design. Instead of focusing on hard-to-solve problems, we advocate for problems characterized by a set of equivalents, equally probable solutions. In this scenario, quantum cryptanalysis is reduced to Grover's scheme, which has exponential complexity in implementation. A classic example of a cryptographic algorithm utilizing a set of possible solutions is the Shamir threshold secret sharing scheme. Shamir's scheme is based on the classical algebra problem of approximating a polynomial by its values. Its secrecy is ensured by the fact that an incomplete number of private secrets (function values) prevents the reconstruction of the overall secret. We have considered logarithmic signature as a basic cryptographic primitive in [8]. Logarithmic signatures possess a strong structural organization due to the necessity of factorizing ciphertexts by the subblocks of the logarithmic signature array. Functionally, logarithmic signature arrays provide a keyless mapping of the input text into ciphertext, essentially functioning as a block cipher in substitution mode for a fixed key or large-scale substitution. The security of the cipher relies on the complexity of decomposing the cryptogram into vectors when the correspondence between vector positions and values is unknown. Quantum cryptanalysis is ineffective if the reverse transformation does not allow for a definitive mapping of the ciphertext back to the input text. If multiple mappings exist relative to the input text, quantum attacks will not succeed. This concept is integral to the cryptosystem being proposed.

The technological advantage of logarithmic signatures lies in the simplicity of computing text ciphers using addition operations with bitwise XOR. However, the disadvantage is the large size of the signature-logarithmic arrays and the need for masking arrays to maintain a high level of secrecy.

## 3  Our contribution

### 3.1  Definition of an incomplete cryptosystem of linear equations

The conceptual property of a digital signature is that it should prevent collisions for various input parameters. This can be achieved by ensuring there is no redundancy of multiple signatures for different input parameters, as seen in block symmetric ciphers, or by imposing specific structural restrictions on the algorithm.

For instance, in the RSA algorithm, this is accomplished through the selection of a pair of keys (characterized by their mutual simplicity relative to the RSA transformation module) and calculations within rings of integers, where structural redundancy manifests as the periodicity of elements within the ring.

Another relevant example is the collision problem in hash schemes. To date, there are no classical solutions for hash functions that guarantee the absence of collisions. Should such solutions exist, they are likely to involve collision scenarios. However, since the sets of input and output parameters are of equal size, hash functions are considered schemes without redundancy. Hash functions exhibit a regular structure and employ cyclic irreversible calculations on consistent parameters. Structurally, hash functions are highly redundant. Unlike classical computers, quantum computers execute reversible calculations, fundamentally violating the condition of irreversibility inherent in hash computations. All NP-complex problems possess structural redundancy by necessity, and cryptosystems built upon them inherently display strong structural redundancy. All current candidates for post-quantum cryptography exhibit significant structural redundancy, prompting efforts to introduce additional constraints and conditions on encryption scheme parameters. Cryptographic algorithms with regularities in their structured data will perpetually face threats from quantum attacks.

To construct a quantum-protected digital signature, we employ an algebraic problem characterized by a set of equivalent, equally probable solutions. This approach shifts from a difficult-to-solve problem to one with multiple solutions, reducing quantum cryptanalysis to Grover's scheme, which exhibits exponential complexity in its implementation. Our cryptosystem is based on a well-known algebraic problem, which asserts that a unique solution exists only for a fully defined system of linear equations. When dealing with an incompletely defined system of equations, the number of solutions corresponds to the size of the set of possible solutions.

We will establish linear equations relative to the unknowns using the values of logarithmic subscripts. The number of equations concerning the secret values of logarithmic signatures is fewer than the number of unknowns. This results in an incomplete system of linear equations, making it impossible to solve in polynomial time. Consequently, the only viable attack on the cryptosystem involves an exhaustive search to define the variables. The security of a cryptosystem built on a problem with incompletely defined equations is determined by the robustness of the set of possible solutions. In this algorithm, the logarithmic signature serves as a fundamental cryptographic primitive, possessing distinct properties such as non-linearity, non-commutability, unidirectionality, and key-dependent factorability.

### 3.2 Our proposal

Let's consider the main steps of the algorithm.

**Step 1. Here we construct of a secret logarithmic signature over a field $F(2^m)$.** The implementation of secret homomorphic transformations with calculations over the field $F(2^m)$ presented in [35]. Let's construct a logarithmic signature using the following set of secret homomorphic transformations:

$$\beta_1 \xrightarrow{\rho_1} \beta_2 \xrightarrow{\rho_2} \beta_3 \xrightarrow{\rho_3} \beta_4 \xrightarrow{\rho_4} \beta_5 \xrightarrow{\rho_5} \beta ,$$

where $\beta_1$ - simple factorization logarithmic signature type $(r_1,...,r_s)_{\beta_1}$;

Transformation 1 ($\rho_1$). In this step we make a noise of $s$ blocks of the $\beta_1$ signature. In this case the signature type does not change. As a result, we get signature $\beta_2$.

Transformation 2 ($\rho_2$). Next, we shuffling secretly all blocks of $\beta_2$ signature. As a result, we get signature $\beta_3$.

Transformation 3 ($\rho_3$). Then, we mix all records in signature blocks $\beta_3$. As a result, we get signature $\beta_4$.

Transformation 4 ($\rho_4$). Next, we proceed with secret homomorphic transformation of array strings in this way: $\beta_4(i) = \gamma \cdot \beta_3(i)$, $i = \overline{1, r_1 + r_2 + ... + r_s}$, $\gamma \in F(2^m)$. As a result, we get signature $\beta_5$.

Transformation 5 ($\rho_5$). Finally, we use secret homomorphic transformation of string array $\beta(i) = \beta_4(i) \cdot \omega_{m \times m}$, $i = \overline{1, r_1 + r_2 + ... + r_s}$. As a result, we get signature $\beta$. Note, that $\omega_{m \times m}$ is an invertible binary matrix of dimension $m \times m$.

We have a logarithmic signature $\beta = \beta(i)$ over $m$-bit strings as a result of all steps. The security estimation is determined taking to the account a high entropy of secret transformations.

The estimates of secrecy are determined by the high entropy of secret transformations and are detailed in [23,33 ÷ 35].

We will fix the downtime logarithmic signature $\beta_1$ with type $(r_1, ..., r_s)_{\beta_1} = (2, ..., 2)_{\beta_1}$, $s = m$. This type provides the lowest costs for the general parameters of the cryptosystem.

Noise $\rho_1$ introduces randomization into records $\beta_1$. The number of noisy bits is determined by the value of $N_1 = \frac{m(m-1)}{2}$. Secrecy has a value $n_1 \approx 0.5m^2$ of bits.

Permutation and merging $\rho_2, \rho_3$ is the next mechanism for ensuring secrecy.

The secret permutation of the signature blocks $\beta_2$ for the type $(2, ..., 2)_{\beta_2}$ is implemented in subsets $\beta_2(1)$ and $\beta_2(2)$, $\beta_2 = \beta_2(1) \| \beta_2(2)$. The subset $\beta_2(1)$ is formed from the first half of blocks $\beta_2$ that contain records with the maximum Hamming weight $\leq m/2$. A subset $\beta_2(1)$ is formed for the second half of the blocks $\beta_2$ with records of maximum weight $> m/2$.

The total number of permutations is equal to $N_2 = ((m/2)!)^2$. Secrecy has a value $n_2 = \log N_2 = 2\log[(m/2)!]$ of bits.

The merging of signature blocks $\beta_3$ is performed according to the fixed rule of merging one block at a time from the first and second sets. Let's get a signature $\beta_4$ with the type $(r_1, ..., r_s)_{\beta_4} = (4, ..., 4)_{\beta_4}$, $s = m/2$.

Permutations of records $\rho_4$ in signature blocks $\beta_4 \to \beta_5$ are independent, their number is equal to $N_3 = (4!)^s = (4!)^{m/2}$. Secrecy has a value $n_3 = \log N_3 = \log(4!)^{m/2} = 0,5m\log(4!) = 2,29m$ of bits.

The secrecy of the homomorphic transformation $\rho_5$ is determined by the size $F(2^m)$ and is equal $n_4 = m$ to bits.

The secrecy of the homomorphic transformation $\rho_6$ is determined by the number of invertible binary matrices $\sigma_{m \times m}$ over $F(2^m)$

$N_5 = (2^m - 1)(2^m - 2)(2^m - 2^2) \cdots (2^m - 2^{m-1})$ and for practically significant ones $m$ has an estimate of $N_5 = 2^{m^2-2}$. Secrecy has a value $n_5 = \log N_5 = m^2 - 2$ of bits.

The secrecy of the homomorphic transformation $\rho_7$ term of the logarithmic signature array $\beta_7$ type $(r_1, ..., r_s)_{\beta_7}$ is determined by the size $F(2^m)$ and level $m$ bit for one signature block i $n_6 = s \cdot m$ bit for the entire signature $\beta$. Since the merging of the signature blocks $\beta_3$ is performed over a pair of blocks $s = m/2$, we will get the $n_6 = 0,5m^2$ bit estimate. The overall rating of secrecy is equal

$$n_{LS} = \sum_{i=1}^{6} n_i = 0,5m^2 + 2\log[(m/2)!] + \log(4!)^{m/2} + m + m^2 - 2 + s \cdot m =$$
$$1,5m^2 + s \cdot m + 2\log[(m/2)!] + \log(4!)^{m/2} + m - 2$$

Secrecy estimates by types of transformations are presented in Table 1.

*Table 1 - Secrecy estimates by types of transformations*

| $m$ | $n_1 \approx 0,5m^2$ beat | $n_2 = 2\log[(m/2)!]$ beat | $n_3 = \log(4!)^{m/2}$ beat | $n_4 = m$ beat | $n_5 = m^2 - 2$ beat | $n_6 = 0,5m^2$ beat | $n_{LS}$ beat |
|---|---|---|---|---|---|---|---|
| 8 | 32 | 9 | 18 | 8 | 62 | 32 | 1 61 |
| 16 | 128 | 30 | 36 | 16 | 254 | 128 | 5 92 |
| 32 | 512 | 88 | 72 | 32 | 1022 | 256 | 1 882 |
| 64 | 2048 | 235 | 144 | 64 | 4094 | 1024 | 7,609 |

Let`s consider attack to the logarithmic signature. The goal of the attack is to determine the desired logarithmic signature $\beta_2$. Conditions of the attack – the attacker knows the general parameters $F(2^m)$, type $(r_1,...,r_s)$, records of the array $\beta$.

Transformations $\rho_1,...,\rho_4$ are not recognized as homomorphic, they introduce uncertainty into the representation of the signature $\beta_5$ and have an entropy $H_1 = n_1 + n_2 + n_3$ of bits. Only one set $\rho_1,...,\rho_4$ is true and unique for creating a signature.

The transformations $\rho_5 \div \rho_7$ are secret homomorphic and perform calculations of the form

$$\beta^{(j)}(i_j) = \gamma \cdot \beta_5^{(j)}(i_j) \cdot \sigma_{m \times m} + t_j, \ i_j = \overline{1, r_j}, \ j = \overline{1, s}, \ t_1,...,t_s \in F(2^m), \ (r_1,...,r_s)_{\beta_7}.$$

Since the records $\beta_5^{(j)}(i_j)$ are not known, we have the task of cryptanalysis of the cipher based on the known ciphertexts.

Here we show possible attack algorithm:
- we generate a transformed $\rho_1,...,\rho_4$ signature with the help of $\beta_5$.
- we solve the problem - for the famous $\beta_5^{(j)}(i_j)$ and $\beta^{(j)}(i_j)$ find conversions $\rho_5 \div \rho_7$.

Complexity rating. Consider equations of the form $\beta_8^{(j)}(i_j) = \gamma \cdot \beta_5^{(j)}(i_j) \cdot \sigma_{m \times m}$.

Such equations can be obtained in case of removal the transformation $\rho_7$ due to the calculation of even sums inside the block. The maximum number of equations is equal to $k_1 = 3m$. The number of unknowns $k_2 = n_4 + n_5 = m^2 + m - 2$. Since $k_1 < k_2$ the system of equations has many solutions. For solving a system of equations, this is a class of complex problems $2^{k_2 - k_1} = 2^{m^2 - 2m - 2}$.

Let us consider equations of the form $\beta^{(j)}(i_j) = \gamma \cdot \beta_5^{(j)}(i_j) \cdot \sigma_{m \times m} + t_j$.

The maximum number of equations is equal to $k_1 = m$.

The number of unknowns $k_2 = n_4 + n_5 + n_6 = 1,5m^2 + m - 2$.

Since $k_1 < k_2$ the system of equations has many solutions.

This is a class of complex problems $2^{k_2-k_1} = 2^{1,5m^2-2}$.

The algorithm stops when $R_i$ a condition is met for a given sample of input parameters $i = \overline{1,s}$

$$\sum_{i=1}^{s} \beta(R_i) = \sum_{i=1}^{s} \beta'(R_i), \text{ where } \beta'_8 \text{ is the required logarithmic signature.}$$

The calculation $\sum_{i=1}^{s} \beta(R_i)$ is determined by the following rule. Let as an argument for $\beta$ is $m$ a bit line $R$. Let's decompose the term $R$ into values according to the type $(r_1, ..., r_s)$

$$R = (R_1, R_2, ..., R_s) = 2^0 R_1 + 2^{\log r_1} R_2 + 2^{\log r_1 r_2} R_3 + ... = R_1 + \sum_{j=2}^{s} 2^{\sum_{i=1}^{j-1} \log r_i} R_j.$$

The values $R_j$ show the number of the record in $j$ the block of the array $\beta$. Calculations for the argument $R$ are determined by bitwise summation of the array terms $\beta$

$$\beta(R) = \beta(R_1, R_2, ..., R_s) = \sum_{j=1}^{s} \beta_{R_j, j}.$$

Check and calculations for $\beta$ is implemented for each signature sample $\beta_5$. The possible number of signatures $\beta_5$ is equal to $2^{n_1+n_2+n_3}$ and the total complexity of the attack is equal to $2^{n_{LS}}$.

The classic application of logarithmic signatures for directional encryption and digital signature [18-23, 32] was based on the problem of the complexity of the inverse transformation $\beta(R)^{-1} = (R_1, R_2, ..., R_s)$ for a large group. In works [24 ÷ 31], logarithmic signatures were proposed for the problem of the complexity of the inverse transformation $\beta(R)^{-1}$ for multiparameter groups with large power over relatively small fields. In all cases, the logarithmic subscript is defined as a large substitution table for mapping the input vector to the output vector. The large uncertainty of reverse transformations, which is included in the estimation of secrecy for calculations in small fields, will not work. It is always possible to construct a substitution table of realizable sizes for small fields.

In our algorithm, we propose a conceptually new idea of using logarithmic signatures - the implementation of cryptographically calculated array terms $\beta_7(i) = \beta_6(i) \cdot \sigma_{m \times m}$, and not at the level of the resulting mapping of the logarithmic signature $\beta(i) = \beta_7^{(j)}(i_j) + t_j$. Scrutiny attacks with substitution of input parameters and analysis of output values are not implemented in this case. The absence of a target function makes Grover's quantum search attack in the database impossible.

**Step 2. Construction of general parameters, public and secret keys of the cryptosystem.**

Let's construct $K = k \times k$ logarithmic signatures $\beta_K$. We present logarithmic signatures $\beta_K$ in the form of a two-dimensional set of arrays with indexing $\beta_{k_1,k_2} \in \beta_K$, $k_1 = \overline{1,k}$, $k_2 = \overline{1,k}$ for given types $r_{k_1,k_2} = (r_1,...,r_{s(k_1,k_2)})_{k_1,k_2}$, $r_{k_1,k_2} \in r_K$.

$\beta_{k_1,k_2} = [B_1, B_2,..., B_{s(k_1,k_2)}]_{k_1,k_2} := (\beta_{i,j})_{k_1,k_2}$, $j = \overline{1, s(k_1,k_2)}$, $i = \overline{1, r_j}$.

The index $j$ determines the number of the block, the index $i$ the number of the record in $j$ the block. Entries of arrays $\beta_{i,j}$ are defined $m$-bitwise terms that we identify with the elements of the finite field $F(2^m)$.

the set of logarithmic signatures $\beta_K$ into two subsets of iz $L$ and $K-L$ logarithmic signatures $\beta_L$, $\beta_{K-L}$,.

Logarithmic signatures $\beta_K$ will be constructed using secret transformations $\rho_1 \div \rho_5$. In two-dimensional indexing $\beta_{k_1,k_2}$, $k_1$ and $k_2$ determine the logarithmic subscript belonging to the set $\beta_{k_1,k_2} \in \beta_K$.

For logarithmic signatures $\beta_K$, we will generate arrays $\alpha_{k_1,k_2} \in \alpha_L$ with random entries $\alpha_{k_1,k_2} = [A_1, A_2,..., A_{s(k_1,k_2)}]_{k_1,k_2} := (a_{i,j})_{k_1,k_2}$, $a_{i,j} \in F(2^m)/0$, $j = \overline{1,s(k_1,k_2)}$, $i = \overline{1, r_j}$, $(r_1,...,r_{s(k_1,k_2)})_{k_1,k_2}$. We will generate random sets

$t_{k_1,k_2} \in t_K$, $t_{k_1,k_2} = (t_1,...,t_{s(k_1,k_2)})_{k_1,k_2} \in F(2^m)/0$

$\tau_{k_1,k_2} \in \tau_K$, $\tau_{k_1,k_2} = (\tau_1,...,\tau_{s(k_1,k_2)})_{k_1,k_2} \in F(2^m)/0$

and let $(t_{i,j})_{k_1,k_2} \neq (\tau_{i,j})_{k_1,k_2}$, $(t_j)_{k_1,k_2} \neq 0$, $(\tau_j)_{k_1,k_2} \neq 0$, $i = \overline{1, r_j}$ - the number of the record in $j = \overline{1, s(k_1,k_2)}$ the block of the array, for a logarithmic signature $\beta_{k_1,k_2}$ of the type $(r_1,...,r_{s(k_1,k_2)})_{k_1,k_2}$. Values $\tau_{k_1,k_2} \in \tau_K$ are recognized as dependent and connected by the system of linear equations presented in the next step.

Let's determine the arrays $\gamma_{k_1,k_2} \in \gamma_K$ and $\lambda_{k_1,k_2} \in \lambda_K$.

For logarithmic subscripts, $\beta_{k_1,k_2} \in \beta_K$ we define arrays $\gamma_{k_1,k_2}$ and $\lambda_{k_1,k_2}$

$(\gamma_{i,j})_{k_1,k_2} = (\beta_{i,j})_{k_1,k_2} + (t_j)_{k_1,k_2} + (\alpha_{i,j})_{k_1,k_2}$,

$(\lambda_{i,j})_{k_1,k_2} = (\alpha_{i,j})_{k_1,k_2} + (\tau_j)_{k_1,k_2}$.

The result is :
- general parameters and cryptosystems $K = k \times k$, $L$, $m$, $r_K$;
- secret keys $\beta_K$, $t_K$, $\tau_K$;
- public keys $\gamma_K$, $\lambda_K$.

**Step 3. Construction of a digital signature based on a cryptosystem with linear equations.**

The digital signature provides proof that the signature associated with the message was calculated on the basis of possession of the secret parameters of a cryptosystem with public keys. A cryptosystem based on a system of linear equations is built on calculations $L$ of linear sums $\sum \gamma_{k_1,k_2}(R_{k_1,k_2}) = U_l$, $l = \overline{1,L}$. We will choose expressions for sums $U_l$ from the following set

$$\sum_{j=1}^{k} \gamma_{ij}(R_{ij}) = U_i, \; i = \overline{1,k}$$

$$\sum_{j=1}^{k} \gamma_{ji}(R_{ji}) = U_{k+i}, \; i = \overline{1,k}$$

$$\sum_{j=1}^{k} \gamma_{j\mu}(R_{j\mu}) = U_{2k+i}, \; \mu = (k-j+i) \bmod k + 1, \; i = \overline{1,k}$$

$$\sum_{j=1}^{k} \gamma_{\mu j}(R_{\mu j}) = U_{3k+i}, \; \mu = (k-i+j) \bmod k + 1, \; i = \overline{1,k}$$

Values $\gamma_{ij}(R_{ij})$ are calculated by $R_{ij}$. Calculations for $\gamma_{k_1,k_2} \in \gamma_K$ and $\lambda_{k_1,k_2} \in \lambda_K$ are determined by the same rules for $\beta$

$$\gamma_{k_1,k_2}(R) = \gamma_{k_1,k_2}(R_1, R_2, ..., R_{s(k_1,k_2)}) = \sum_{j=1}^{s(k_1,k_2)} \gamma_{R_j,j}.$$

Expressions for $U_l$ selected from consideration that include one value each $\gamma_{k_1,k_2}(R_{k_1,k_2})$ from a row and/or column of the array $\gamma_{k_1,k_2} \in \gamma_K$. Such will be expressed $4k$. Relatively, $\gamma_{ij}(R_{ij})$ we get a system of linear equations. Since the number of unknowns $\gamma_{k_1,k_2}(R_{k_1,k_2})$ is equal $K = k^2$, and the number of knowns $U_l$ is equal $L < K$, the system of linear equations will be incomplete with respect to the unknowns $\gamma_{k_1,k_2}(R_{k_1,k_2})$. Relatively well-known $U_l$, $l = \overline{1,L}$, values $\gamma_{k_1,k_2}(R_{k_1,k_2})$ are recognized as related. For $K$ values of logarithmic signatures, $\gamma_{k_1,k_2}(R_{k_1,k_2})$ it is easy to calculate $L < K$ the values of $U_l$. The solution of the inverse problem regarding the finding $\gamma_{k_1,k_2}(R_{k_1,k_2})$ has uncertainty among $2^{(K-L)m}$ the possible solutions.

**Figure 1 - Arrays**

The cryptosystem has potential $(K - L)m$ bit secrecy. Example Let $k = 4$ In Fig. 1, arrays are marked in brown $\gamma_{k_1,k_2}$, which define expressions for $U_i$ $i = \overline{1,4k}$.

For construction cryptosystems with secrecy $(K - L)m$ beat we will form $L$ equations $U_L$ that are linearly independent relative to the desired ones $\gamma_{k_1,k_2} \in \gamma_L$.

Let $L = 8$. Let's choose the following eight equations $U_L = \{U_1, U_2, U_3, U_5, U_9 \div U_{12}\}$. Expressions for relatively unknown amounts $\gamma_{k_1,k_2} \in \gamma_K$ have the following form

$$\gamma_{11}(R_{11}) + \gamma_{21}(R_{21}) + \gamma_{31}(R_{31}) + \gamma_{41}(R_{41}) = U_1$$
$$\gamma_{12}(R_{12}) + \gamma_{22}(R_{22}) + \gamma_{32}(R_{32}) + \gamma_{42}(R_{42}) = U_2$$
$$\gamma_{13}(R_{13}) + \gamma_{23}(R_{23}) + \gamma_{33}(R_{33}) + \gamma_{43}(R_{43}) = U_3$$
$$\gamma_{11}(R_{11}) + \gamma_{12}(R_{12}) + \gamma_{13}(R_{13}) + \gamma_{14}(R_{14}) = U_5$$
$$\gamma_{11}(R_{11}) + \gamma_{24}(R_{24}) + \gamma_{33}(R_{33}) + \gamma_{42}(R_{42}) = U_9$$
$$\gamma_{12}(R_{12}) + \gamma_{21}(R_{21}) + \gamma_{34}(R_{34}) + \gamma_{43}(R_{43}) = U_{10}$$
$$\gamma_{13}(R_{13}) + \gamma_{22}(R_{22}) + \gamma_{31}(R_{31}) + \gamma_{44}(R_{44}) = U_{11}$$
$$\gamma_{14}(R_{14}) + \gamma_{23}(R_{23}) + \gamma_{32}(R_{32}) + \gamma_{41}(R_{41}) = U_{12}$$

The solution for the unknowns $\gamma_{k_1,k_2} \in \gamma_L$ can be expressed in the following expressions

$$\gamma_{11}(R_{11}) = U_9 + \gamma_{24}(R_{24}) + \gamma_{33}(R_{33}) + \gamma_{42}(R_{42})$$
$$\gamma_{12}(R_{12}) = (U_3 + U_5 + U_9 + U_{12}) + \gamma_{24}(R_{24}) + \gamma_{32}(R_{32}) + \gamma_{41}(R_{41}) + \gamma_{42}(R_{42}) + \gamma_{43}(R_{43})$$
$$\gamma_{13}(R_{13}) = (U_1 + U_2 + U_9 + U_{10} + U_{11}) + \gamma_{24}(R_{24}) + \gamma_{32}(R_{32})$$
$$+ \gamma_{33}(R_{33}) + \gamma_{34}(R_{34}) + \gamma_{41}(R_{41}) + \gamma_{43}(R_{43}) + \gamma_{44}(R_{44})$$
$$\gamma_{14}(R_{14}) = (U_1 + U_2 + U_3 + U_9 + U_{10} + U_{11} + U_{12}) + \gamma_{24}(R_{24}) + \gamma_{34}(R_{34}) + \gamma_{44}(R_{44})$$
$$\gamma_{21}(R_{21}) = (U_3 + U_5 + U_9 + U_{10} + U_{12}) + \gamma_{24}(R_{24}) + \gamma_{32}(R_{32}) + \gamma_{34}(R_{34}) + \gamma_{41}(R_{41}) + \gamma_{42}(R_{42})$$
$$\gamma_{22}(R_{22}) = (U_2 + U_3 + U_5 + U_9 + U_{12}) + \gamma_{24}(R_{24}) + \gamma_{41}(R_{41}) + \gamma_{43}(R_{43})$$
$$\gamma_{23}(R_{23}) = (U_1 + U_2 + U_3 + U_9 + U_{10} + U_{11}) + \gamma_{24}(R_{24}) + \gamma_{32}(R_{32}) + \gamma_{34}(R_{34}) + \gamma_{41}(R_{41}) + \gamma_{44}(R_{44})$$
$$\gamma_{31}(R_{31}) = (U_1 + U_3 + U_5 + U_{10} + U_{12}) + \gamma_{32}(R_{32}) + \gamma_{33}(R_{33}) + \gamma_{34}(R_{34})$$

Thus, to calculate the values, $\{\gamma_{11}, \gamma_{12}, \gamma_{13}, \gamma_{14}, \gamma_{21}, \gamma_{22}, \gamma_{23}, \gamma_{31}\} \in \gamma_L$ one should define $\{\gamma_{24}, \gamma_{32}, \gamma_{33}, \gamma_{34}, \gamma_{41}, \gamma_{42}, \gamma_{43}, \gamma_{44}\} \in G_{K-L}$.

**Step 4. Calculation of a digital signature.**

Input parameters: message $x$, secret keys $\beta_K$, $t_K$ hash function $h$.

Calculation of a digital signature consists of the following steps.

1) We will generate a session key $\upsilon$ with the size $L \cdot m$ of bits and calculate the hash value $h(x, \upsilon)$ for the message $x$ and the key $\upsilon$. The hash function $h$ is unidirectional and sensitive to bit changes in the message $x$ and session key $\upsilon$. Let's introduce $h(x,\upsilon)$ $K-L$ $m$- in bit terms $h(x,\upsilon) = h(x,\upsilon)_1 \| h(x,\upsilon)_2 \| ... \| h(x,\upsilon)_{K-L}$.

2) For the input values $R_l = h(x,\upsilon)_l$, $l = \overline{1, K-L}$ with reindexing on the parameters, $k_1, k_2$ we will calculate $G_{k_1,k_2}(R_{k_1,k_2}) = \beta_{k_1,k_2}(R_{k_1,k_2}) + t_{k_1,k_2}(R_{k_1,k_2})$, $\beta_{k_1,k_2} \in \beta_{K-L}$,

3) Let's present the session key $\upsilon$ $m$- in bit terms $\upsilon = \upsilon_1 \| \upsilon_2 \| ... \| \upsilon_L$ and let $U_l = \upsilon_l$, $l = \overline{1, L}$.

4) Let's construct $L$ linear equations for $U_l \in U_L$ relative $G_{k_1,k_2}(R_{k_1,k_2})$ to similar expressions $\sum \gamma_{k_1,k_2}(R_{k_1,k_2}) = U_l$

$\sum G_{k_1,k_2}(R_{k_1,k_2}) = U_l$, $l = \overline{1, L}$, $\beta_{k_1,k_2} \in \beta_K$.

6) Let's solve the equations with respect to the unknowns $G_{k_1,k_2}(R_{k_1,k_2})$, $\beta_{k_1,k_2} \in \beta_L$. Let's find it $R_{k_1,k_2}$ through factorization $\beta_{k_1,k_2} \in \beta_L$ for $G_{k_1,k_2}(R_{k_1,k_2})$

$$R_{k_1,k_2} = \beta_{k_1,k_2}^{-1}(\sum_{\gamma_{k_1,k_2} \in \gamma_L} G_{k_1,k_2}(R_{k_1,k_2}) + t_{k_1,k_2}) = \beta_{k_1,k_2}^{-1}(\sum_{\gamma_{k_1,k_2} \in \gamma_{K-L}} G_{k_1,k_2}(R_{k_1,k_2}) + \sum_{l \in \overline{1,L}} U_l + t_{k_1,k_2})$$

,

The result of the digital signature for the message $x$ is a tuple of values $R_{k_1,k_2}$, $k_1 = \overline{1, k}$, $k_2 = \overline{1, k}$.

**Step 5. Verification of digital signature**

Input parameters message $x$, $R_{k_1,k_2}$, $k_1 = \overline{1, k}$, $k_2 = \overline{1, k}$, public keys $\gamma_K$, $\lambda_K$, hash function $h$. Verification of a digital signature consists of the following steps.

1) Calculate $D_l = W_l + V_l$, $l = \overline{1, L}$, where

$W_l = \sum_{\gamma_{k_1,k_2} \in \gamma_K} \gamma_{k_1,k_2}(R_{k_1,k_2})$,

$V_l = \sum_{\lambda_{k_1,k_2} \in \lambda_K} \lambda_{k_1,k_2}(R_{k_1,k_2})$.

Indices $k_1 = \varsigma(j,l)$ are $k_2 = \varsigma(j,l)$ determined by the serial number $j$ of the logarithmic signature in the equations for $W_l$ and the number of the equation $l$.

Let's substitute $W_l$ and $V_l$ in the expression for $D_l$, we get

$$D_l = \sum_{\gamma_{k_1,k_2} \in \gamma_K} \gamma_{k_1,k_2}(R_{k_1,k_2}) + \sum_{\lambda_{k_1,k_2} \in \lambda_K} \lambda_{k_1,k_2}(R_{k_1,k_2}) =$$

$$\left( \sum_{\gamma_{k_1,k_2} \in \gamma_L} \beta_{k_1,k_2}(R_{k_1,k_2}) + \sum_{\gamma_{k_1,k_2} \in \gamma_L} t_{k_1,k_2} + \sum_{\gamma_{k_1,k_2} \in \gamma_{k-L}} \beta_{k_1,k_2}(R_{k_1,k_2}) + \sum_{\gamma_{k_1,k_2} \in \gamma_{K-L}} t_{k_1,k_2} + \sum_{\gamma_{k_1,k_2} \in \gamma_K} \alpha_{k_1,k_2}(R_{k_1,k_2}) \right) +$$

$$\left( \sum_{\lambda_{k_1,k_2} \in \lambda_K} \alpha_{k_1,k_2}(R_{k_1,k_2}) + \sum_{k_1,k_2 \in \lambda_K} \tau_{k_1,k_2} \right)$$

$$= \sum_{\gamma_{k_1,k_2} \in \gamma_L} (G_{k_1,k_2}(R_{k_1,k_2}) + t_{k_1,k_2}) + \sum_{\gamma_{k_1,k_2} \in \gamma_L} t_{k_1,k_2} + \sum_{\gamma_{k_1,k_2} \in \gamma_{k-L}} G_{k_1,k_2}(R_{k_1,k_2}) + \sum_{k_1,k_2 \in \lambda_K} \tau_{k_1,k_2}$$

$$= U_l + \sum_{\gamma_{k_1,k_2} \in \gamma_{K-L}} G_{k_1,k_2}(R_{k_1,k_2}) + \sum_{\gamma_{k_1,k_2} \in \gamma_L} t_{k_1,k_2} + \sum_{\gamma_{k_1,k_2} \in \gamma_L} t_{k_1,k_2} + \sum_{\gamma_{k_1,k_2} \in \gamma_{k-L}} G_{k_1,k_2}(R_{k_1,k_2}) + \sum_{k_1,k_2 \in \lambda_K} \tau_{k_1,k_2} = U_l'.$$

The derivation is expressed for $D_l = W_l + V_l$ to the values $U_l'$ is realized by fulfilling the conditions $\sum_{k_1,k_2 \in \lambda_K} \tau_{k_1,k_2} = 0$ for the values $\tau_{k_1,k_2}$. The values $\tau_{k_1,k_2}$ are secret parameters, and they are determined through a system of equations for $L$ linear sums analogous to the expressions $\sum \gamma_{k_1,k_2}(R_{k_1,k_2}) = U_l$, $l = \overline{1,L}$. This condition is necessary for verification of the digital signature and was stipulated at the time of generation $\tau_{k_1,k_2}$.

The system of $L$ equations $\sum_{k_1,k_2 \in K} \tau_{k_1,k_2} = 0$ leads $L < K$ to $2^{K-L}$ possible solutions for $\tau_{k_1,k_2}$.

2) Let's introduce $\upsilon' = U_1' \| U_2' \| \ldots \| U_L'$. Let's calculate the hash value $h(x, \upsilon')$ for the message $x$ and the key $\upsilon'$. Let's introduce $h(x, \upsilon') = h(x, \upsilon')_1 \| h(x, \upsilon')_2 \| \ldots \| h(x, \upsilon')_{K-L}$ $K - L$ $m$ - in bit terms.

3) For the input values $R_l' = h(x, \upsilon')_l$, $l = \overline{1, K-L}$ with parameter reindexing, $k_1, k_2$ we will perform a comparison for equality $R_{k_1,k_2} = R_{k_1,k_2}'$, by indexes $k_1, k_2$ which determine whether the logarithmic signature belongs to the set $\beta_{k_1,k_2} \in \beta_K$.

If there is a tie $R_{k_1,k_2} = R_{k_1,k_2}'$, the digital signature is confirmed.

### 3.3 Security analysis

Let us consider attacks on a digital signature in a cryptosystem with linear equations.

The goal of the attack is to forge a digital signature for a message $x$. We consider a following attack algorithm.

1) We generate the session key $\upsilon$ with the size $L \cdot m$ of bits and calculate the hash value $h(x, \upsilon)$.

2) We obtain the values of $R_l = h(x, \upsilon)_l$, $l = \overline{1, K-L}$, through the representation of $h(x, \upsilon) = h(x, \upsilon)_1 \| h(x, \upsilon)_2 \| \ldots \| h(x, \upsilon)_{K-L}$. After reindexing by parameters, $k_1, k_2$ we can calculate $K - L$ the vectors $\gamma_{k_1,k_2}(R_{k_1,k_2})$ and $\lambda_{k_1,k_2}(R_{k_1,k_2})$.

The problem of cryptanalysis. Based on the values of $\gamma_{k_1,k_2}(R_{k_1,k_2})$ and $\lambda_{k_1,k_2}(R_{k_1,k_2})$, $\beta_{k_1,k_2} \in \beta_{K-L}$ it is impossible to calculate y $G_{k_1,k_2}(R_{k_1,k_2}) = \beta_{k_1,k_2}(R_{k_1,k_2}) + t_{k_1,k_2}(R_{k_1,k_2})$ and make up the equations $\sum G_{k_1,k_2}(R_{k_1,k_2}) = U_l$, $l = \overline{1,L}$, $\beta_{k_1,k_2} \in \beta_K$.

Addition $\gamma_{k_1,k_2}(R_{k_1,k_2})$ and $\lambda_{k_1,k_2}(R_{k_1,k_2})$ leads to the result

$$\gamma_{k_1,k_2}(R_{k_1,k_2}) + \lambda_{k_1,k_2}(R_{k_1,k_2}) = \beta_{k_1,k_2}(R_{k_1,k_2}) + t_{k_1,k_2}(R_{k_1,k_2}) + \tau_{k_1,k_2}(R_{k_1,k_2}) = G_{k_1,k_2}(R_{k_1,k_2}) + \tau_{k_1,k_2}(R_{k_1,k_2})$$

.

From these, it is impossible to obtain the calculated values $G_{k_1,k_2}(R_{k_1,k_2})$, $\beta_{k_1,k_2} \in \beta_{K-L}$ and it is impossible to obtain a solution for the unknowns $G_{k_1,k_2}(R_{k_1,k_2})$, $\beta_{k_1,k_2} \in \beta_L$.

3) We have only an attack of a full search by $R_{k_1,k_2}$. Values $R_l = h(x,\upsilon)_l$, $l = \overline{1, K-L}$ recognized by the session key and $L$ values $R_{k_1,k_2}$ for $\beta_{k_1,k_2} \in \beta_L$ are defined by a search until the result is achieved

$$D_l = \sum_{\gamma_{k_1,k_2} \in \gamma_K} \gamma_{k_1,k_2}(R_{k_1,k_2}) + \sum_{\lambda_{k_1,k_2} \in \lambda_K} \lambda_{k_1,k_2}(R_{k_1,k_2}),$$

$D_l = \upsilon_l$.

C falsity of the attack $N_1 = 2^{Lm}$.

4) Evaluation of the quantum secrecy of a digital signature based on a cryptosystem with an incomplete system of linear equations.

### 3.4 Grover's quantum algorithm with exponential complexity.

Here we consider estimates of secrecy and implementation costs.
The general parameters of the cryptosystem are recognized
- $m$ bit length of logarithmic signatures;
- $K$ the number of logarithmic signatures in the cryptosystem;
- $L$ the number of equations in the cryptosystem;
- $r_K$ types of logarithmic signatures.

The assessment of the secrecy of a digital signature for a brute-force attack is determined by the number of linear equations $L$ on a set of $K$ logarithmic signatures. In our construction, we used $4k$ equations and in the maximum set $L_{max} = 4k$ we can specify $L < L_{max}$ non-degenerate equations. For $k = 4$ it was obtained that $L \leq 11$ non-degenerate equations can be selected. With sufficient confidence, it is possible to assume that such equations exist $L = 2k$. Possible parameters of the cryptosystem for various $m$, $K$, $L$ and secrecy ratings are presented in the table 2.

*Table 2 - Parameters*

| $k$ | $K$ | $L$ | Attack brute force (blow) | | | |
|---|---|---|---|---|---|---|
| | | | $m = 8$ | $m = 16$ | $m = 32$ | $m = 64$ |
| 4 | 16 | 8 | 64 | 128 | 256 | 512 |
| 6 | 24 | 12 | 96 | 192 | 384 | 768 |
| 8 | 32 | 16 | 128 | 256 | 512 | 1024 |

The sizes of keys for building a cryptosystem for various $m$, $K$, $L$ and let are presented below $r_{k_1,k_2} = \overbrace{(2,2,...,2)}^{m}$. Arrays $\lambda_K$ can be generated as random entries based on the initial value. The secret keys of the cryptosystem are $\beta_K$, $t_K$, $\tau_K$. Public keys are recognized as $\gamma_K$, $\lambda_K$.

*Table 3 - Costs for secret keys*

| $m$ | $|\beta_K| = 2Km$ byte | | | $|t_K| = |\tau_K| = Km$ byte | | | $|\beta_K| + |t_K| + |\tau_K|$ byte | | | Secrecy (beat) | | |
|---|---|---|---|---|---|---|---|---|---|---|---|---|
| K= | 16 | 24 | 32 | 16 | 24 | 32 | 16 | 24 | 32 | 16 | 24 | 32 |
| 8 | 256 | 384 | 512 | 128 | 192 | 256 | 512 | 768 | 1024 | 64 | 96 | 128 |
| 16 | 512 | 768 | 1024 | 256 | 384 | 512 | 1024 | 1536 | 2048 | 128 | 192 | 256 |
| 32 | 1024 | 1536 | 2048 | 512 | 768 | 1024 | 2048 | 3072 | 4096 | 256 | 384 | 512 |
| 64 | 2048 | 3072 | 4096 | 1024 | 1536 | 2048 | 4096 | 6144 | 8192 | 512 | 768 | 1024 |

*Table 4 - Public key costs*

| m | $|\gamma_K| = |\beta_K|$ byte | | | $|\lambda_K| = |\alpha_K|$ byte | | | $|\gamma_K| + |\lambda_K|$ byte | | |
|---|---|---|---|---|---|---|---|---|---|
| K= | 16 | 24 | 32 | 16 | 24 | 32 | 16 | 24 | 32 |
| 8 | 256 | 384 | 512 | 256 | 384 | 512 | 512 | 768 | 1024 |
| 16 | 512 | 768 | 1024 | 512 | 768 | 1024 | 1024 | 1536 | 2048 |
| 32 | 1024 | 1536 | 2048 | 1024 | 1536 | 2048 | 2048 | 3072 | 4096 |
| 64 | 2048 | 3072 | 4096 | 2048 | 3072 | 4096 | 4096 | 6144 | 8192 |

*Table 5 - Digital signature costs*

| m | The size of the digital signature $L \cdot m$ (beat) | | | The number is added during calculation $G_{k_1,k_2}(R_{k_1,k_2})$ $L(K-L)/2$ | | | Number of factorizations $G_{k_1,k_2}(R_{k_1,k_2})$ $L$ | | |
|---|---|---|---|---|---|---|---|---|---|
| | K=16 L=8 | K=24 L=12 | K=32 L=16 | K=16 L=8 | K=24 L=12 | K=32 L=16 | K=16 L=8 | K=24 L=12 | K=32 L=16 |
| 8 | 64 | 96 | 128 | 32 | 72 | 128 | 8 | 12 | 16 |
| 16 | 128 | 192 | 256 | 32 | 72 | 128 | 8 | 12 | 16 |
| 32 | 256 | 384 | 512 | 32 | 72 | 128 | 8 | 12 | 16 |
| 64 | 512 | 768 | 1024 | 32 | 72 | 128 | 8 | 12 | 16 |

*Table 6 - Digital signature verification costs*

| m | The number is added during calculation $W_l = \sum_{\gamma_{k_1,k_2} \in \gamma_K} \gamma_{k_1,k_2}(R_{k_1,k_2})$ $K \cdot m$ | | | The number is added during calculation $V_l = \sum_{\lambda_{k_1,k_2} \in \lambda_K} \lambda_{k_1,k_2}(R_{k_1,k_2})$ $K \cdot m$ | | | The total number is compounded $W_l + V_l$ $2K \cdot m$ | | |
|---|---|---|---|---|---|---|---|---|---|
| | K=16 L=8 | K=24 L=12 | K=32 L=16 | K=16 L=8 | K=24 L=12 | K=32 L=16 | K=16 L=8 | K=24 L=12 | K=32 L=16 |
| 8 | 128 | 192 | 256 | 128 | 192 | 256 | 256 | 384 | 512 |
| 16 | 256 | 384 | 512 | 256 | 384 | 512 | 512 | 768 | 1024 |
| 32 | 512 | 768 | 1024 | 512 | 768 | 1024 | 1024 | 1536 | 2048 |
| 64 | 024 | 1536 | 2048 | 1024 | 1536 | 2048 | 2048 | 3072 | 4096 |

To calculate the digital signature, it is necessary to compose $L$ linear equations with respect to $G_{k_1,k_2}(R_{k_1,k_2})$, $\beta_{k_1,k_2} \in \beta_L$, calculate them and find them $R_{k_1,k_2}$ through factorization $\beta_{k_1,k_2} \in \beta_L$ for $G_{k_1,k_2}(R_{k_1,k_2})$. Estimates of costs for calculating a digital signature and checking with $m$ bit words are presented in the tables 4 and 5.

We do not take into account hashing costs $h(x,\upsilon)$ for the message $x$ and key $\upsilon$ when calculating the digital signature and verification. These costs are traditional for any algorithm.

## 4  Conclusions

A cryptosystem based on a system of linear equations with logarithmic signatures is an excellent candidate for post-quantum cryptography. Its quantum security is derived from the impossibility of establishing a correspondence between the input and output of the logarithmic signature. Bridging attacks, which involve substituting input parameters and analyzing output values, are infeasible due to the absence of a target function. This characteristic renders Grover's quantum search attack ineffective. However, the cryptosystem still faces a formidable brute-force attack through the substitution of digital signature parameters.

The public key sizes for the declared NIST security levels of 128, 192, and 256 bits are 1 KB, 1.5 KB, and 2 KB, respectively. These sizes are comparable to the best current candidates for post-quantum cryptography. The digital signature algorithm is highly scalable in terms of computational costs, memory requirements, and hardware platform limitations, without compromising its high level of security. The basic operation of the algorithm involves bitwise XOR over words of 8, 16, 32, and 64 bits of logarithmic arrays. Furthermore, implementing higher levels of secrecy, measured in bits, is straightforward with keys of appropriate sizes in kilobytes. A cryptosystem based on linear equations guarantees the indecipherability of secret logarithmic signatures.